# Sense me: Supporting awareness in parent-child relationships through mobile sensing

**José Rodrigues, Rúben Gouveia, Olga Lyra, Evangelos Karapanos**
Madeira Interactive Technologies Institute, Funchal, Portugal
{javr.rodrigues, rubahfgouveia, lyra.olga, e.karapanos}@gmail.com

## ABSTRACT
We introduce Sense$_\mu$ (pronounced "sense me"), a mobile application that aims at supporting awareness in parent-child relationships through the sensing capabilities of mobile devices. We discuss the relevance of three types of awareness information: *physical activity* inferred from accelerometers, *verbal activity* during class hours inferred from microphones, and *social activity* inferred from Bluetooth pair-wise proximity sensing. We describe how we attempt to contextualize these sensing data with the goal of supporting parents' awareness of the educational performance and social wellbeing of their children, as well as motivating and sustaining a two-way communication between parents and teachers over the long term.

## INTRODUCTION
Awareness systems are increasingly becoming an alternative communication aid as they support sustained and effortless communication based on a peripheral awareness of the activities and the status of the parties involved [1]. Earlier studies have shown that awareness systems have the capacity to address a variety of intra-family communication needs ranging from instrumental coordination needs to affective needs of connectedness, companionship and reassurance (c.f. [1]).

With the decreasing available time for communication in dual-income and distributed families, increased interest has recently been paid to the communication needs between parents and children.

Khan et al. [4], for instance, developed KidzFrame, a distributed application that enables daycare professionals to take photos of children and communicate these to parents along with short messages. Brown et al. [2] developed the whereabouts clock, an awareness system designed that displays the current location of family members, thus providing a feeling of reassurance that everything is according to schedule. Yarosh et al. [10] developed ShareTable, a camera-projector system that supports synchronous interactions between parents and children in different locations. ShareTable allows the sharing of a view of physical artifacts through projections over a surface, thus supporting non-verbal communication between parents and children. Another relevant work is the digital family portrait [8], a concept of a digital family portrait, which populates itself with useful information about the conditions of family members that live in a different household.

A particular challenge in the design of awareness system relates to the distinction between automatically sensed awareness cues and ones being deliberately captured and communicated by the user. While automatically sensed awareness cues have the capacity to establish and maintain a continuous awareness of each other's lives thus affecting long-term psychological states such as a feeling of companionship [9], such cues need to be sensitive to the particular context in order to support meaningful inferences by the recipient [4].

## SENSEME
Our goal in the design and development of Sense$_\mu$ was to create a system that requires minimum input from the child, yet supports meaningful inferences about the child's educational performance and social wellbeing. To achieve this we support a two-way communication between the teacher and the parent over sensed data, motivated by recent evidence on the limited school-home communication [11]. Sense$_\mu$ attempts to infer three indicators of educational performance and social wellbeing:

*Physical activity* during breaks is inferred from the accelerometer with a one-sec sampling rate. While previous research has focused on inferring type of physical activity such as walking or running [8], our goal was to establish an overall index of physical activity and compare this to the mean value of a one-week history. Sampling is restricted to the scheduled periods of school breaks.

*Verbal activity* during class hours is inferred from the microphone – "one of the most ubiquitous and unexploited sensors in mobile phones" [7] – with a one-sec sampling. An overall index of verbal activity for each class is compared against the mean value of a one-week history of the same class.

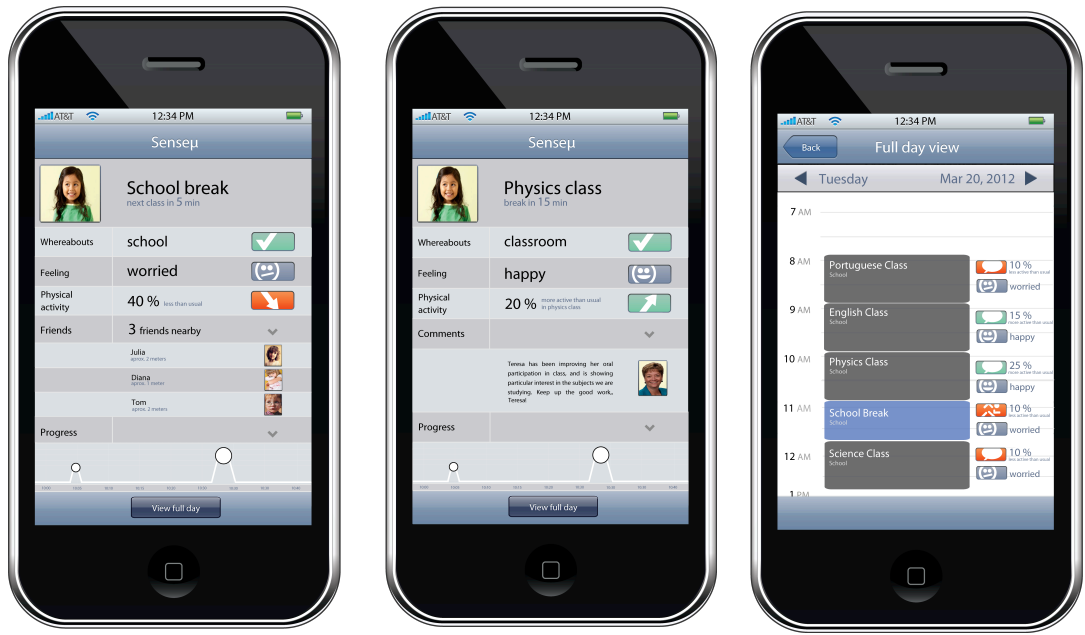

*Social Activity* is inferred from pair-wise proximity sensing using Bluetooth technology. Bluetooth has been widely used in measure social activity, recognizing user's social patterns, relationships, and modeling organizational rhythms.

Beyond the automatic capture of awareness cues, Sense$_\mu$ enables children to self-report affective information using a predefined set of icons and emotion labels. Next, a web application allows the teacher to review all physical, verbal, and social activity cues, and annotate them. Communication between the teacher and the parent is two-directional with the goal of motivating and sustaining communication over the long run [11]. Having children' privacy in mind, Sense$_\mu$ is designed to support inferences from minimal awareness cues, it attempts to obfuscate sensitive information (e.g. presenting abstract location such as "at school" and "in class" rather than exact location), and does not record conversations or store precise data (e.g., exact locations).

**ONGOING STUDIES**
To inform the design of Sense$_\mu$ we conducted a Day Reconstruction study followed by contextual inquiry with 17 families. on their daily communication activities and awareness needs. A feasibility study of Sense$_\mu$ is currently planned with the aim of assessing the validity of sensed data as indicators of verbal, physical and social activity, as well as uncover unexpected problems (e.g., changes in schedule). Next, a long-term deployment of Sense$_\mu$ will attempt to inquire into how families adopt it and appropriate it in daily life [12] as well inquire into the systems' impact on parents' daily awareness of their kids' activities, educational performance and social wellbeing, as well as home-school communication.